\documentclass[12pt,preprint]{aastex}

\shorttitle{L Dwarf RV}
\shortauthors{Blake  et al.}

\begin{document}

\title{Multiepoch Radial Velocity Observations of L Dwarfs}

\author{Cullen H. Blake\altaffilmark{1}, David Charbonneau\altaffilmark{2}}
\affil{Harvard-Smithsonian Center for Astrophysics, 60 Garden Street, Cambridge, MA 02138; cblake@cfa.harvard.edu}

\author{Russel  J. White}
\affil{University of Alabama in Huntsville, Physics Department, 301 Sparkman Drive, 201B Optics Building, Huntsville, AL 35899}

\author{Mark S. Marley}
\affil{NASA Ames Research Center, MS 245-3, Moffett Field, CA 94035 }

\author{Didier Saumon}
\affil{Los Alamos National Laboratory, MS F663, Los Alamos, NM 87545 }

\altaffiltext{1}{Harvard Origins of Life Initiative Fellow}
\altaffiltext{2}{Alfred P. Sloan Research Fellow}

\begin{abstract}
We report on the development of a technique for precise radial-velocity measurements of cool stars and brown dwarfs in the near infrared. Our technique is analogous to the Iodine ($I_{2}$) absorption cell method that has proven so successful in the optical regime. We rely on telluric CH$_{4}$ absorption features to serve as a wavelength reference, relative to which we measure  Doppler shifts of the CO and H$_{2}$O features in the spectra of our targets.  We apply this technique to high-resolution (R$\approx$50,000) spectra near 2.3$\micron$ of nine L dwarfs taken with the Phoenix instrument on Gemini-South and demonstrate a typical precision of 300 m s$^{-1}$. We conduct simulations to estimate our expected precision and show our performance is currently limited by the signal-to-noise of our data. We present estimates of the rotational velocities and systemic velocities of our targets. With our current data, we are sensitive to companions with $M \sin i>2M_{\rm{J}}$ in orbits with periods less than three days.  We identify no companions in our current data set. Future observations with improved signal-to-noise should result in radial-velocity precision of 100 m s$^{-1}$ for L dwarfs.
\end{abstract}

\keywords{stars: low-mass, brown dwarfs; techniques: radial velocity; planetary systems}

 \section{Introduction}

 The majority of planet searches focus on sun-like stars, dwarfs of late K to early F spectral type.  While stars smaller than these are far more common  (e.g. Henry et al. 2006), extension of the well-developed radial-velocity techniques  to these cooler, dimmer targets poses many difficulties. Searching for planetary companions to stars with masses a fraction of that of the sun presents the obvious advantage that at equal orbital periods,  companions of smaller masses can be detected with a given radial velocity precision. The two most successful radial-velocity techniques, the $I_{2}$ absorption cell and Th-Ar comparison lamp methods, have both been applied to stars of early M spectral type. This has resulted in the discovery of four M-dwarf planetary systems: GJ 876 \citep{delfosse1998,rivera2005}, GJ 436 \citep{butler2004}, GJ 581 \citep{bonfils2005}, and GJ 674 \citep{bonfils2007}.  \citet{endl2003,endl2006}  carried out a systematic search of 90 M dwarfs, most of spectral type earlier than M3. These authors achieved an average precision of 8 m s$^{-1}$ and estimated that the rate of Jupiter-mass companions within $a<0.7$AU of these objects is less than $1.3\%$, a result marginally inconsistent with radial-velocity observations of larger main sequence stars \citep{marcy2005, udry2007}.  
 
\citet{ida2005} have conducted theoretical simulations of planet formation via core accretion, and predict  that companions of Neptune-mass and smaller should be common in short-period orbits around lower main-sequence stars. However, their simulations do not extend down to the lowest mass stars, or to brown dwarfs. Work by both \citet{laughlin2004} and \citet{boss2006} specifically addresses the formation of planets around low-mass stars. They also predict that Neptune-mass companions should be relatively common, while Jupiter-mass companions should be much less common orbiting low-mass stars than sun-like stars.  The microlensing detection of two sub-Neptune mass planets orbiting M dwarfs tentatively supports this hypothesis \citep{gould2006,beaulieu2006}. Searching for giant companions to low-mass stars represents an important test of current planet formation theories. 
 
Today we know of a large number of objects populating the lowest reaches of the main sequence (e.g. Cruz et al. 2003a) as well as objects too low in mass to reach the main sequence. These objects, classified into the late M, L and T spectral types, have masses close to the minimum mass required for main sequence hydrogen burning. Objects in the L  class straddle the $0.075M_{\sun}$ limit that is commonly invoked to delineate brown dwarfs from stars \citep{burrows2001}. The question of the formation of companions to these very low-mass stars has been little explored. There is, however, strong evidence from observations of infrared excesses of very low-mass stars in star-forming regions (see Luhman et al. 2007  for a review) that, when young, these objects possess protoplanetary disks at a rate similar to that of more massive stars.  
 
 At fixed orbital period, the reflex radial velocity amplitude due to a companion of a given mass scales with host mass as $M_{\star}^{-2/3}$, making an L dwarf with a mass of $0.1M_{\sun} = 105 M_{\rm{J}}$ a very accessible target for a radial-velocity search for planetary companions. Specifically, the semi-amplitude of the radial-velocity signal from a planet of mass M$_{\rm{p}}$ orbiting an L dwarf with a period $P$, orbital inclination $i$, and eccentricity $e$ is:
 
\begin{equation}
K  \approx 654\ {\rm{m}\ \rm{s}^{-1}} \left(P\over{3\rm{d}}\right)^{-1/3} \left({M_{\rm{p}}}\over{M_{\rm{J}}}\right)  \left({M_{\star}}\over{0.1M_{\sun}}\right)^{-2/3}  \left(1-e^{2}\right)^{1/2}  \sin{i}
\end{equation}

The masses of field L dwarfs are not yet well established from purely dynamical estimates. \citet{burrows2001} find that the mass of a low-mass star may be estimated from its age $t$ and effective temperature $T_{\rm{eff}}$ according to

\begin{equation}
{{M_{\star}}\over{M_{\sun}}} \approx   0.05  \left({T_{\rm{eff}}}\over{1550\rm{K}}\right)^{1.2} \left({{1\rm{Gyr}}\over{t}}\right)^{-0.38}.
\end{equation}
If we  assume that all of our objects are older than 1Gyr and have $T_{\rm{eff}}$ in the range 1800K to 2400K \citep{dahn2002}  we infer that all of our targets have a mass of approximately $0.1M_{\sun}$ since mass depends rather weakly on $T_{\rm{eff}}$ and age for our field (presumably old) objects. This may overestimate the mass of some of our targets by 15$\%$ or more, but our ability to detect planets only improves with lower primary mass, hence this is a conservative assumption. A detection of the resonance absorption line of Li  in the spectrum of a low-mass star would constrain the mass to be less than 65 M$_{\rm{J}}$ \citep{magazzu1993,kirkpatrick1999,martin1999}. A literature search identified no searches for Li in any of our targets. Assuming $M_\star = 0.1 M_\Sun$, a companion with $M_{p} \sin{i} = M_{\rm{J}}$ with an orbital period of 3d would induce a radial-velocity variation with a peak-to-peak amplitude of 1.3 km s$^{-1}$. By optical radial velocity standards this is an enormous signal, but precision sufficient for the detection of this signal has never been demonstrated in observations of L dwarfs.  

The extremely low luminosity of these objects makes obtaining precise radial velocities difficult.  \citet{guenther2003}  observed one L dwarf in the optical using Th-Ar comparison lamps as a wavelength reference and obtained a precision of 1.3 km s$^{-1}$. \citet{bailer-jones2004} observed a sample of early L dwarfs in the optical and used a cross-correlation technique to produce radial-velocity measurements at a single epoch with a precision between 0.5 and 4 km s$^{-1}$. The extremely red colors of these objects ($V-K$=10.0 at L0; Dahn et al. 2002) motivate a move to the near infrared (NIR) and the development of a set of techniques that could be used to produce precise radial velocities with spectrographs operating in that wavelength range. \citet{martin2006} present observations of the young brown dwarf LP 944-20 (spectral type M9) near $1.25 \micron$ using Keck/NIRSPEC and report a precision between 0.4 and 1.6 km s$^{-1}$ from their cross correlation technique. These authors also present optical measurements of LP 944-20, some in common with \citet{guenther2003}, and find that their NIR techniques are achieving similar precision to the optical techniques on this object. The optical radial velocities of LP 944-20 exhibit significant variation while the NIR radial velocities do not. \citet{martin2006} suggest that spots, with greater contrast in the optical than the NIR, are responsible for the observed radial velocity variations and conclude that radial-velocity measurements of low-mass stars in the optical may be fundamentally limited by the stars themselves. 

Here we describe a technique for high-precision radial-velocity measurements of very low-mass stars in the NIR and present some initial results. This technique, which uses the telluric absorption features of the Earth's atmosphere in place of a gas  absorption cell, exploits the rich radial velocity content of the CO bandhead in the $K$-band spectra of these cool objects. In Section \ref{observations}, we describe our observations and the extraction of the spectra. In Section \ref{modeling}, we describe our modeling procedure, and in Section \ref{res} we present results from a single observing run spanning five nights, and place limits on the presence of orbiting companions for our targets.

\section{Observations} \label{observations}

The use of  telluric absorption features as a reference from which to derive precise radial velocities of stars has a long history in the literature. \citet{balthasar1982}, \citet{smith1982}, and \citet{caccin1985}  demonstrated the use of telluric O$_{2}$ with a stability of 10 m s$^{-1}$. The NIR region is rich with telluric features due to CH$_{4}$. The L dwarfs contain a wide array of their own absorption features due to molecules such as H$_{2}$O  and CO. At these effective temperatures, the dominant carbon-bearing molecule is CO, not CH$_{4}$, so there is no confusion between object and telluric CH$_{4}$.  The bandhead due to the $2-0$ transition of CO at 2.3 $\micron$ results in a set of periodic features in the L dwarf spectrum. We seek to exploit the superposition of the telluric CH$_{4}$ and L dwarf CO and H$_{2}$O features in order to extract precise radial velocities from high-resolution observations  at these wavelengths.

We observed a sample of nine L dwarfs with the Phoenix spectrograph  \citep{hinkle2003}  on Gemini-South over five nights spanning UT 2006 January 10-14 under observing program GS-2005B-C-3. Observing conditions were mostly free of clouds during this observing run, and seeing remained between 0.3\arcsec and 0.8\arcsec in K band. We observed all of our targets at airmass between 1.0 and 1.6. We used a slit with a width of 3 pixels (0.26\arcsec) to yield an approximate resolution of R=50,000 with the single-order spectra spanning $0.012\micron$, from $2.296\micron$ to $2.308\micron$. Exposure times ranged from 1000 to 1800 seconds per nod position and were scaled so that the expected S/N per pixel in the extracted spectra would be approximately 10.0.  In practice we found that the S/N of our many of our extracted spectra fell below this expectation by up to $50\%$. This is primarily due to guiding-induced jitter coupled to slit losses as Phoenix has no slit-viewing camera. We gathered observations in nodded pairs, with consecutive exposures offset in opposite directions relative to the center of the slit. This observing strategy facilitates subtraction of the bright sky lines.  Our targets, described in Table 1, range in spectral type from L0 and L6 and were selected from the sample of known L dwarfs brighter than K=12.  
We observed each object at least three times during the observing run and observed seven of the nine L dwarfs four times. We also observed rapidly rotating A stars each night in order to help calibrate the spectrograph PSF from the telluric lines superimposed upon the featureless spectra of these hot stars.

We created a single flat field for each night by first subtracting a median dark image of the same exposure time from each of a nightly set of 15 individual internal flats and then using a median combination of normalized frames to create a final flat field image. After dividing each science frame by the flat field image, the sky lines and dark current were effectively removed from the 2-D spectra by subtracting nod pairs. We extracted the 1-D spectra from the sky-subtracted nod pairs following the optimal extraction procedure outlined by \citet{horne1986}.  We created a bad pixel mask through visual inspection of the flat field images in order to mask bad pixels during optimal extraction procedure. We  propagated the flux of the sky lines from the individual images in order to correctly evaluate the noise estimate on the flux extracted from the subtracted nod pairs. We excised  200 pixels from the blue end of each spectrum that were contaminated by cosmetic defects and a strong sensitivity gradient in that region of the detector. The final 1-D spectra spanned 800 pixels centered around the wavelength $2.303\micron$. Representative spectra are plotted in Figures 1 through 3.

\section{Spectral Modeling and Fitting} \label{modeling}
We estimated the radial velocities of the targets using a detailed modeling procedure.  Our methods are similar to those used in the $I_{2}$ cell method \citep{butler1996}. Following a strategy similar to that described in \citet{valenti1995}, we build up a many-parameter model to describe the spectrum of our target, the spectrograph response, and the Earth's atmosphere. 

Our model begins with two high-resolution template spectra. A high-resolution ($5\times10^{-6} \micron$ pixel$^{-1}$) spectrum of the Earth's atmosphere is provided by \citet{livingston1776}. This spectrum 
is derived from Fourier transform interferometer observations of the Sun at different airmasses.  We also use a high-resolution synthetic spectrum  of an L dwarf prepared specifically for this project. These synthetic  spectra are computed as described in \citet{marley2002}, with updated opacities
as described in \citet{freedman2007}.  However, only the line opacities  of $\rm CH_4$, CO, $\rm H_2O$, TiO and $\rm H_2S$ are relevant in this  wavelength region. Continuum opacities include the pressure-induced opacity of molecular hydrogen and helium and the Mie opacity of the iron and silicate clouds.
To treat the clouds, the models apply the condensation cloud model of \citet{ackerman2001} with a  sedimentation parameter of $f_{\rm sed}=3$, corresponding to a moderate amount of  condensate settling. Both absorption and non-isotropic scattering from the clouds are accounted 
for in the radiative transfer. The models used here have solar metallicity \citep{lodders2003} and a fixed gravity of $\log g=5$ (cgs) and cover a range of $T_{\rm eff}$ from 1200 to 2400$\,$K. Complete details of the models will be presented in an upcoming publication.  Fits of model spectra to moderate resolution data in the near-infrared are presented in \citet{cushing2007}. The synthetic spectra provide monochromatic fluxes spaced $4.2 \times  10^{-6}\,\mu$m apart. We convolve and re-sample the product of these high-resolution spectra to generate the model that we then fit to the extracted 1-D spectra. Our model has several free parameters. The parameters related to the L dwarf are $V\sin i$, $T_{\rm{eff}}$, and radial velocity and the parameters related to the spectrograph are the  PSF
width, flux normalization, and  the wavelength solution (i.e. the  mapping from wavelength to pixel position).

We find that a simple Gaussian model of the PSF  is sufficient for our data and additional parameters do not significantly improve the resulting fits. We also find no evidence for variation of the PSF width across the spectrograph order. Such a simple model likely works well because the spectrograph is not cross dispersed, spans only a very small range of wavelength, and we have relatively few ($\approx10$) telluric lines, each of low S/N. We describe  the wavelength-to-pixel mapping with a second-order polynomial function. We experimented with higher-order functions but found that the additional parameters did not improve the quality of the resulting fits. There is a strong throughput gradient in the blue-most $\approx100$ pixels of all of our spectra. Presumably due to an inadequate calibration, this was easily corrected by a two-parameter normalization process: the first parameter determines the overall level of the spectrum while the second parameter describes a linear, sloping normalization applied to the first 100 pixels.
 
 The first step in our model generating procedure is to Doppler shift the wavelengths corresponding to each monochromatic intensity in the synthetic L dwarf spectrum.  The synthetic L dwarf and telluric spectra are then super-sampled using cubic-spline interpolation onto a regular grid in wavelength at a resolution of approximately $1\times10^{-6}\micron$ pixel$^{-1}$, about 10 times the resolution of the observed spectra.  We experimented with linear and quadratic methods of interpolation and found only small differences in the cores of absorption lines, which did not alter our final results. The L dwarf spectrum is then convolved with a rotational broadening kernel following \citet{gray1992}. We assume no limb darkening in the rotational broadening kernel. It is necessary to scale the telluric spectrum from airmass 2.0 to the airmass of each observation. We use a logarithmic scaling relation empirically determined from Phoenix observations of rapidly rotating A stars at a range of airmasses. The synthetic L dwarf and telluric spectra are multiplied together and this product is convolved with a Gaussian representation of the PSF of the spectrograph. The basic steps in the creation of the super-resolution model spectrum $M(\lambda)$ can be expressed 
 
 \begin{equation}
 M(\lambda) = \left(\left[L\left(\lambda \times \left(1+{v\over{c}}\right)\right)\star  K\right] \times T(\lambda)\right) \star PSF
\end{equation}
 where $\star$ indicates convolution, $L(\lambda)$ is the synthetic L dwarf spectrum, $K$ is the rotational broadening kernel, $T(\lambda)$ is the telluric spectrum, and $PSF$ is the spectrograph point spread function. 
 
We resample this super-resolution spectrum, which is defined in terms of wavelength, onto the detector grid, which is defined in terms of pixel number.  This mapping is described by a second-order polynomial. We experimented with two methods of accomplishing this task. The first involved integrating the finely-sampled spectra over the extent of each detector pixel. This produced nearly identical results to a cubic spline interpolation for estimating the value at the detector position defined by the center of each pixel. We chose to use the cubic spline method for mapping the spectrum onto the pixel grid. Lastly, the entire model spectrum, now at the same resolution as the actual data, is multiplied by the two normalization factors described earlier.  The steps in our procedure are illustrated in Figure 1, which shows the model spectra at each stage in the modeling process. 

In order to estimate the parameters we calculate the goodness of fit between our model and the data and produce a $\chi^{2}$ value. We then utilize the  \textit{amoeba} downhill-simplex minimization procedure \citep{press1986} to  minimize $\chi^{2}$ and estimate the best-fit parameters for each of our spectra.  For each object we executed our fitting scheme in two phases. First, the $V\sin i$ value was allowed to vary and model L dwarf spectra of the 13 different effective temperatures were used to generate our model. We assigned to each object  a $T_{\rm{eff}}$  corresponding to the theoretical L dwarf spectrum that yielded the lowest $\chi^2$ value. This value is not meant to be an estimate of the actual $T_{\rm{eff}}$. Using this best-fit $T_{\rm{eff}}$ the $\chi^2$ surface in $V\sin i$ was determined for each  observed spectrum by fitting the data with a range of fixed $V\sin i$ values while allowing the other parameters to float. A parabolic fit to this surface produced a $V\sin i$ estimate for each observed spectrum for each object as well as an error estimate. We used the weighted average of the $V\sin i$ estimates from each spectrum to estimate the $V\sin i$ and an error set to the rms of the individual estimates.  This statistical $V\sin i$ error likely underestimates the true error on our determination of this parameter. Experiments with theoretical  L dwarf spectra with surface gravities other than $\log g=5$ indicated that the $V\sin i$ values we estimate depend on the surface gravity of the model. Models with $\log g=5.5$ resulted in $V\sin i$ estimates that were systematically lower by $1-2$ km s$^{-1}$, albeit with fits of significantly lower quality. This first phase of the fitting process assumes no degeneracy between $V\sin i$ and $T_{\rm{eff}}$. We performed tests with artificial data designed to mimic the properties of our Phoenix observations (see below) and found this assumption to be valid.  In the second phase of our fitting process we fix the values of $T_{\rm{eff}}$ and $V\sin i$ and use \textit{amoeba} to  determine the seven remaining parameters. These seven parameters are: wavelength solution (3), flux normalization (2), RV(1), PSF width (1). The best-fit radial-velocity estimates are corrected to the solar system barycenter using the IRAF routine \textit{bcvcorr} (G. Torres; private communication). Observed spectra and best-fit models for  two of our targets are shown in Figures 2 and 3. 

It is not trivial to assess the accuracy with which these parameters are determined 
when using a minimization scheme like \textit{amoeba}.  We conducted simulations in an attempt to estimate the errors on our radial velocity determinations as a function of S/N and $V\sin i$.  We first estimated the S/N of the extracted spectra by taking the median of the S/N values of the 800 pixels. We then simulated Phoenix spectra of L dwarfs following the prescription above but with the additional step of adding noise, determined by photon statistics and the gain and readnoise of the Phoenix  detector \citep{hinkle2003},  following the final re-sampling onto the pixel grid. These simulated Phoenix spectra were generated with spectrograph-related (PSF, wavelength solution) parameters fixed to the averages of the values determined from the data, at an airmass of 1.2, and radial velocity drawn from a uniform distribution between $-$20 and 20 km s$^{-1}$. We fed these simulated spectra into our \textit{amoeba} parameter estimation code  to estimate the accuracy with which the known L dwarf parameters could then be recovered. Here, we assumed the correct input template but experiments with different templates found that changing the template had little effect on the derived parameters, with the exception of template spectra with $T_{\rm{eff}}<1600$K (below that of any of our targets) .  We carried out the simulation over a grid of S/N from 2 to 100 and $V\sin i$ from 5 to 50 km s$^{-1}$. The results of this simulation are shown in Figure 4. Comparison of the scatter of the radial-velocity estimates of our targets to the single measurement errors  predicted by our simulations indicates that we are achieving the expected precision. We therefore assign errors to our radial-velocity measurements based on interpolations of the results of our simulations. 

\section{Results} \label{res}
For each of our targets we find rotation velocities in excess of the 6 km s$^{-1}$ velocity resolution of our spectrograph. Like \citet{bailer-jones2004} and \citet{mohanty2003} we find that our L dwarfs are relatively rapid rotators compared to field M dwarfs \citep{delfosse1998b}. We estimate $V\sin i$ values between 7.06 and 20.68 km s$^{-1}$ for our nine targets. We present the best fit $V\sin i$ values and systemic velocities ($V_{\gamma}$) for each of our targets in Table 1 .   The errors on $V\sin i$ are derived from the fits described in Section \ref{modeling} and are based on the standard deviation of the $V\sin i$ values from the individual observations. The value for $V_{\gamma}$ is the average of the individual radial-velocity measurements, and the error we quote is the average error divided by $N^{1/2}$, where $N$ is the number of individual observations. The individual measured radial velocities are presented in Table 2 and the radial velocity curves for each target are shown in Figures 5,6, and 7. For six of our targets we achieve a  radial velocity precision less than 450 m s$^{-1}$ per individual measurement. With our  typical S/N  between 5 and 15, it is clear from Figure 4 that we are currently limited by the noise of our observations and not by systematic effects. Increasing our S/N to 20 would improve our radial velocity precision to 150m s$^{-1}$. As a result, avenues for future improvement in our precision are promising. 

None of the objects show evidence for radial velocity variability. In each case the fit to the data of the null hypothesis of constant radial velocity is good with $\chi_{\nu}^2 \approx 1.0$.  There is evidence that the error estimates derived from our Monte-Carlo simulations are modestly overestimated, as three of our targets have $\chi_{\nu}^2\le0.5$. The source of this discrepancy is not immediately clear. An incorrect estimate of either $V\sin i$ or the median S/N of the 800 pixels in the spectra could lead to an incorrect estimate of the errors. We note, however, that these low $\chi^{2}$ values are not unexpected considering the small number of observations of each object and the total number of objects observed. Our lowest $\chi^{2}$ value (2M1048+01, $\chi_{\nu}^{2}=0.34$) has a probability of $6.4\%$, which is not surprising.

While we don't have a sufficient number of targets in our current sample to place interesting limits on the frequency of close-in, Jupiter-mass companions to L dwarfs, we can quantify the masses of companions that we would have securely detected in the observations of our nine targets.  We carried out such a simulation by injecting sinusoidal radial-velocity signals into the data for each target and examining the statistics of the resulting radial-velocity curves. For each target we created $10^{3}$ fake data sets by the addition of a signal due to a planetary companion with $0.5<M \sin i <10 M_{\rm{J}}$, a period $1.0<P<3.0$ days, and random orbital phase. For each simulation we assume an L dwarf mass of 100 $M_{\rm{J}}$ and estimate the value of $M \sin i$ at which we would be $95\%$ likely to detect a deviation in $\chi^2$ with a confidence of $95\%$. We find that for all of our targets  we can effectively rule out companions of $M \sin i > 4.5 M_{\rm{J}}$ with periods of less than 3 days and that for four of our targets (2M0109$-$51, 2M0835$-$08, 2M0355+11, 2M1045$-$01) companions  with $M \sin i \ge 2.0 M_{J}$  can be excluded in that period range. 

\section{Conclusions}

We describe a method to estimate the multiepoch radial velocities of L dwarfs with a precision sufficient to detect short-period Jupiter-mass companions. Our technique uses atmospheric telluric CH$_{4}$ absorption features as the wavelength reference, and the molecular CO and H$_{2}$O absorption spectrum in the L dwarf to estimate the radial-velocity offset. We have demonstrated an average  precision of ~300 m s$^{-1}$ over a time span of five nights. Monte-Carlo simulations show that even a modest increases in the S/N of our data should lead to a factor of 3 improvement in our precision.  This level of precision opens up the possibility of detecting Jupiter-mass planets with orbital periods of  months.  Improvements in S/N and radial-velocity precision can be realized with current and future higher-throughput instruments on large telescopes.  An understanding of the frequency of planetary companions to objects at the bottom of the main sequence will provide important insight into the process of planet formation.  Moreover, the indirect detection of such planets orbiting nearby low-mass stars and brown dwarfs affords the opportunity for direct study of the planetary emission and characterization of the atmosphere. The ratio of the infrared brightness of a giant planet to its host star (smaller than $10^{-5}$ at $\lambda<10\micron$ for a giant planet orbiting sun-like star at 0.5AU; Burrows et al. 2004) is improved by several orders of magnitude when the host star is an L dwarf.  If it is possible to overcome the overall faintness of the system, direct detection of the planet through high-resolution imaging \citep{beuzit2007} may be possible. Furthermore, if the planet is found to transit its host, observations of secondary eclipses \citep{deming2005, deming2006, charbonneau2006} will result in direct measurements of the planet and its properties.

\acknowledgments

We would like the thank an anonymous referee whose comments helped improve our manuscript. We thank Ken Hinkle and Verne Smith for assistance with the Phoenix spectrograph. The authors would like to thank David Latham and Willie Torres for helpful discussions that contributed to this work.  C.H.B would like to acknowledge support from the Harvard Origins of Life Initiative. This research has benefitted from the M, L, and T dwarf compendium housed at DwarfArchives.org and maintained by Chris Gelino, Davy Kirkpatrick, and Adam Burgasser.  This publication makes use of data products from the Two Micron All Sky Survey, which is a joint project of the University of Massachusetts and
the Infrared Processing and Analysis Center/California Institute of Technology, funded by the National Aeronautics and Space Administration and the National Science Foundation. This work is based on observations obtained at the Gemini Observatory, which is operated by the Association of Universities for Research in Astronomy, Inc., under a cooperative agreement with the NSF on behalf of the Gemini partnership: the National Science Foundation (United States), the Particle Physics and Astronomy Research Council (United Kingdom), the National Research Council (Canada), CONICYT (Chile), the Australian Research Council (Australia), CNPq (Brazil) and CONICET (Argentina). This paper is based on observations obtained with the Phoenix infrared spectrograph, developed and operated by the National Optical Astronomy Observatory.

{\it Facilities:} \facility{Gemini-South/Phoenix}

\clearpage

\begin{figure}
\epsscale{1.0}
\plotone{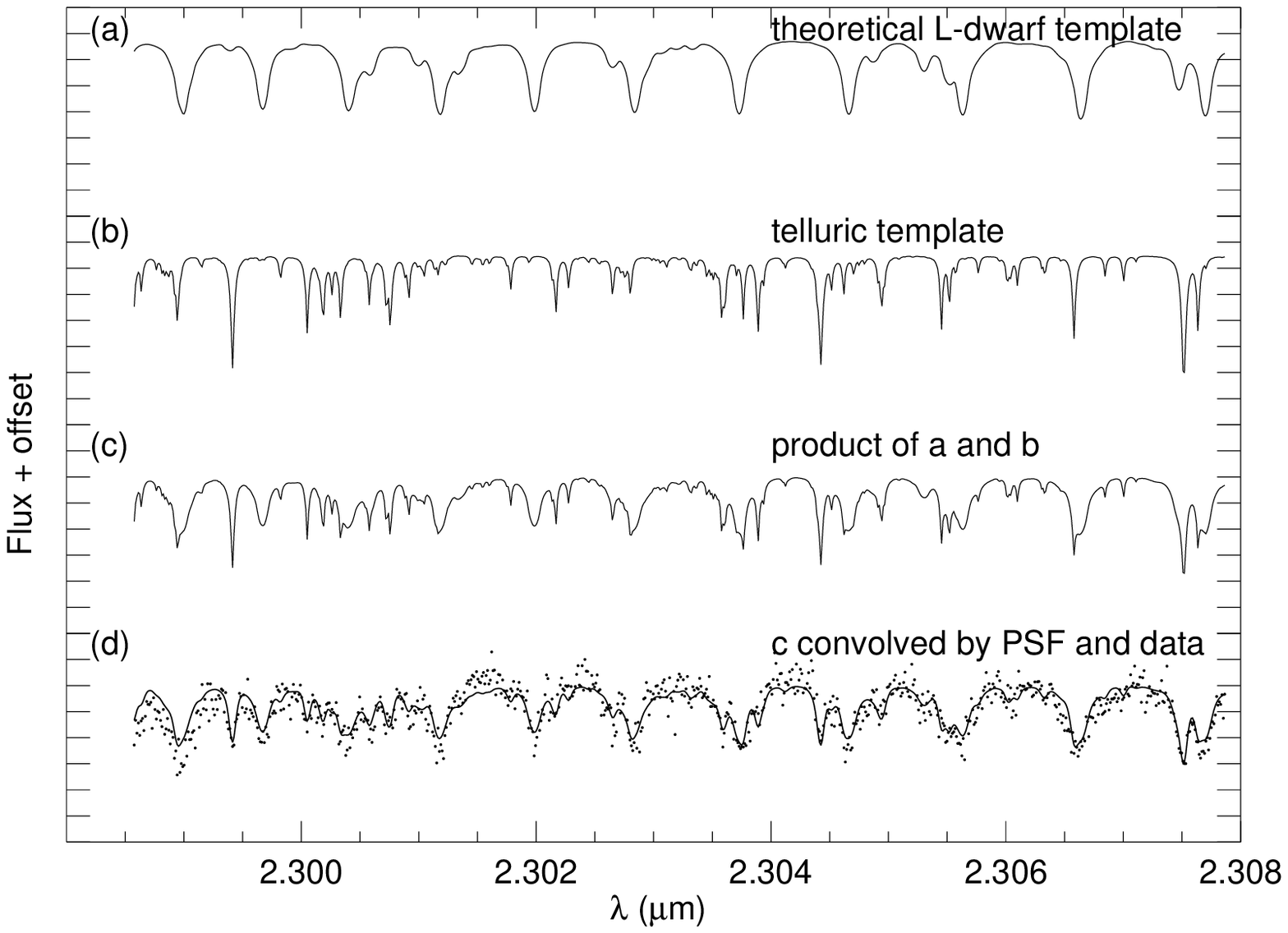}
\caption{Illustrative example of the steps in our modeling process: (a) is the rotationally broadened L dwarf theoretical spectrum, (b) is the observed KPNO telluric spectrum described by \citet{livingston1776},  (c) is the product of (a) and (b),  and (d) is the spectrum (c) convolved with the  PSF. 
The data (small points) in spectrum (d) are the observed spectrum from our Phoenix observations of 2M1155-37. }
\end{figure}

\begin{figure}
\epsscale{1.0}
\plotone{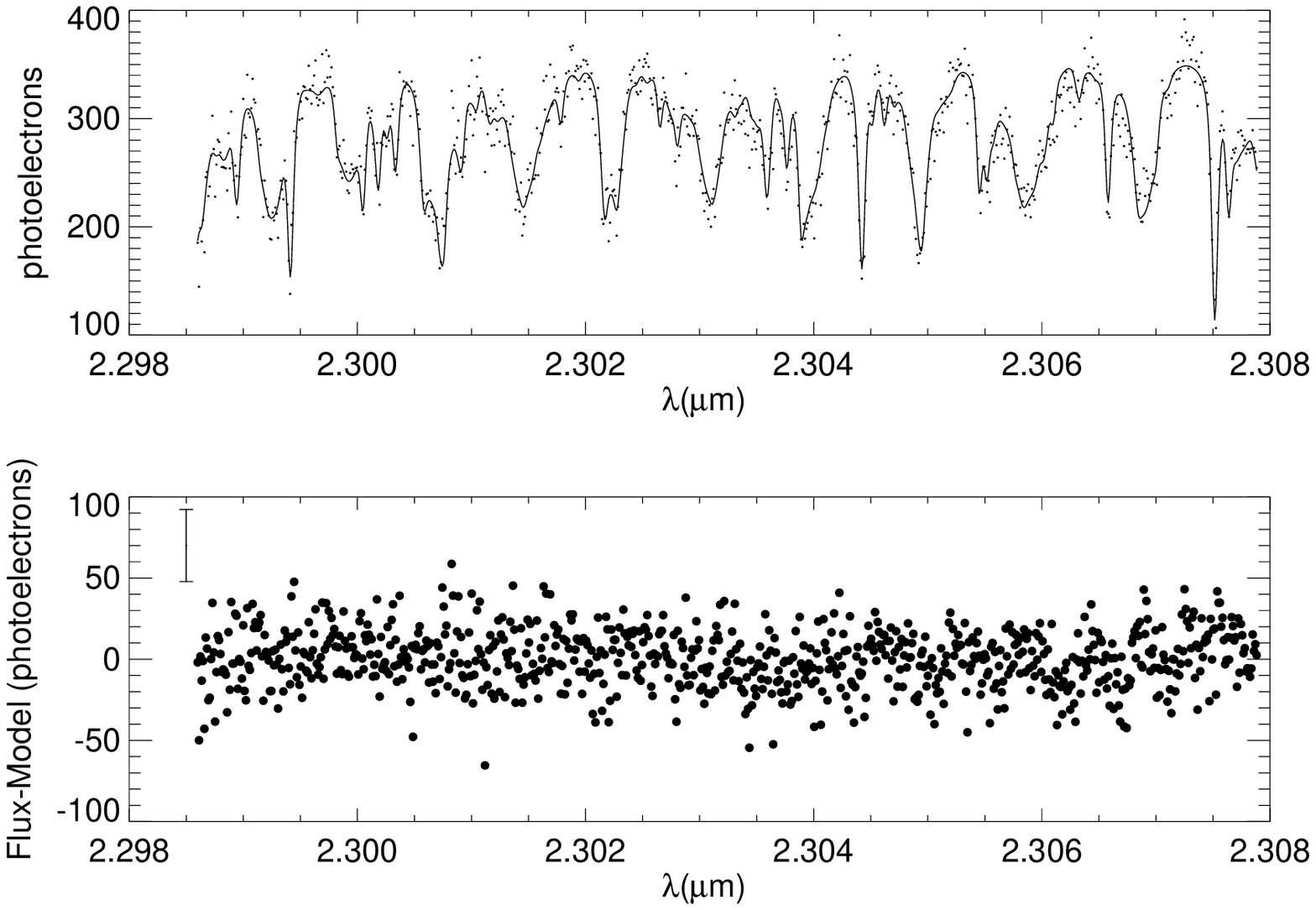}
\caption{Example of a model fit to an observed spectrum of 2M0109-51. The top panel shows the observed data as small points with the best-fit model (determined by the \textit{amoeba} fitting procedure) overplotted as a line. The bottom panel shows the residuals of this fit. The error bar in the upper left corner approximates the expected noise (photon and read noise) of the spectra. We estimate the $V\sin i$ of 2M0109-51 to be 14.58$\pm$0.52 km s$^{-1}$. }
\end{figure}

\begin{figure}
\epsscale{1.0}
\plotone{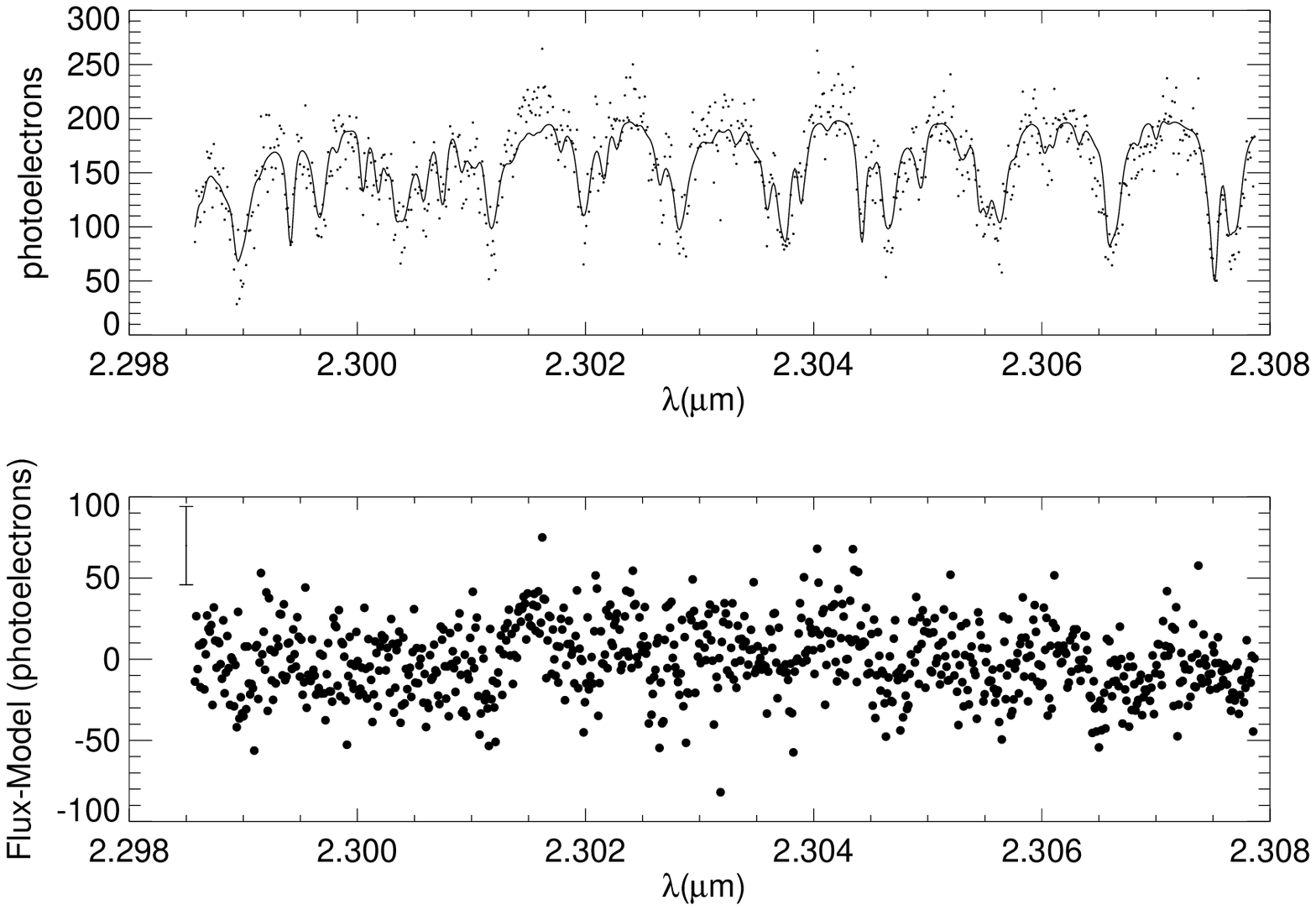}
\caption{Example of a model fit to an observed spectrum of 2M1045-01. The top panel shows the observed data as small points with the best-fit model (determined by the \textit{amoeba} fitting procedure) overplotted as a line. The bottom panel shows the residuals of this fit. The error bar in the upper left corner approximates the expected noise (photon and read noise) of the spectra. We estimate the $V\sin i$ of 2M1045-01 to be 7.06$\pm$0.71 km s$^{-1}$. }
\end{figure}

\begin{figure}
\epsscale{1.0}
\plotone{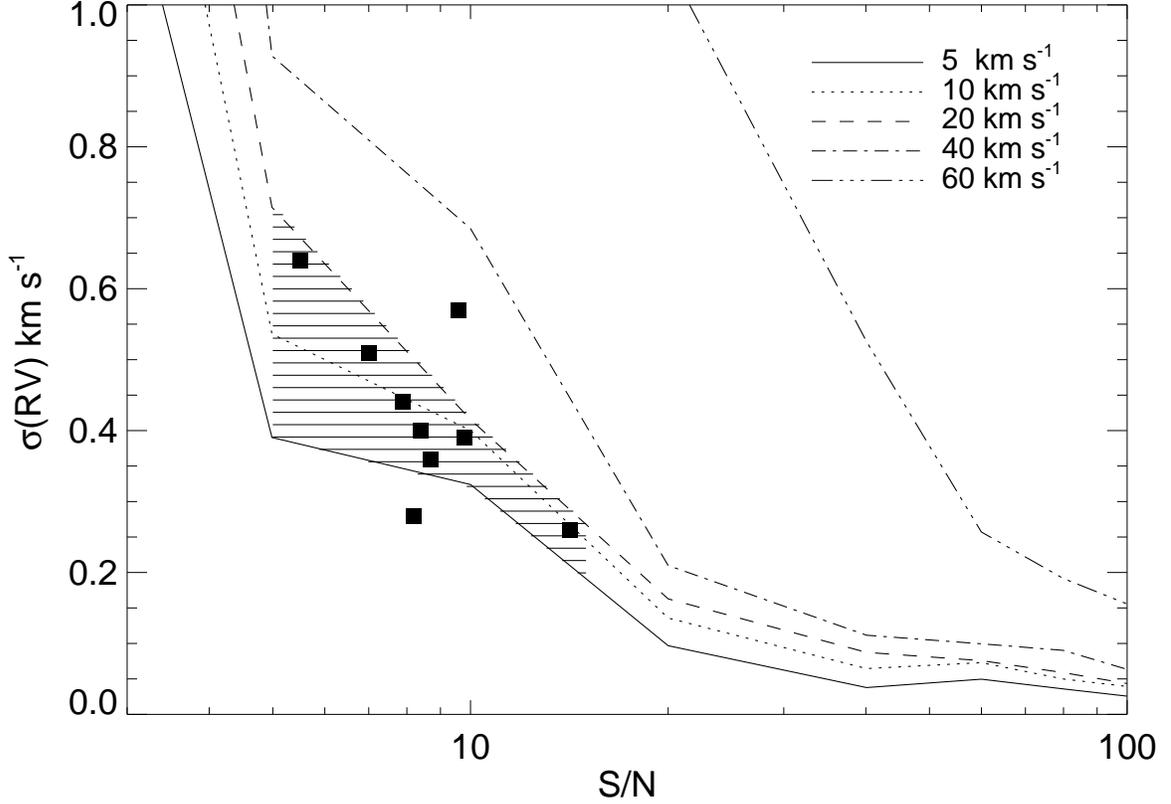}
\caption{Results from our Monte-Carlo simulation to estimate the radial-velocity precision as a function of median S/N per pixel and $V\sin i$. Each line represents a different  $V\sin i$ value in km s$^{-1}$.  Our data have S/N between 5 and 15 and $V\sin i$ between 7 and 21 km s$^{-1}$, corresponding to the shaded region of the plot. The RMS of the radial velocities of each target, along with the average S/N of the observations, are plotted as solid squares. In general, our achieved precision follows the expectations from our simulations. The RMS of the observations of 2M0109$-$51 is about 150 m s$^{-1}$ larger than expected. This large RMS results from the pair of observations taken at the third epoch (see Figure 5).} 
\end{figure}

\begin{figure}
\epsscale{1.0}
\plotone{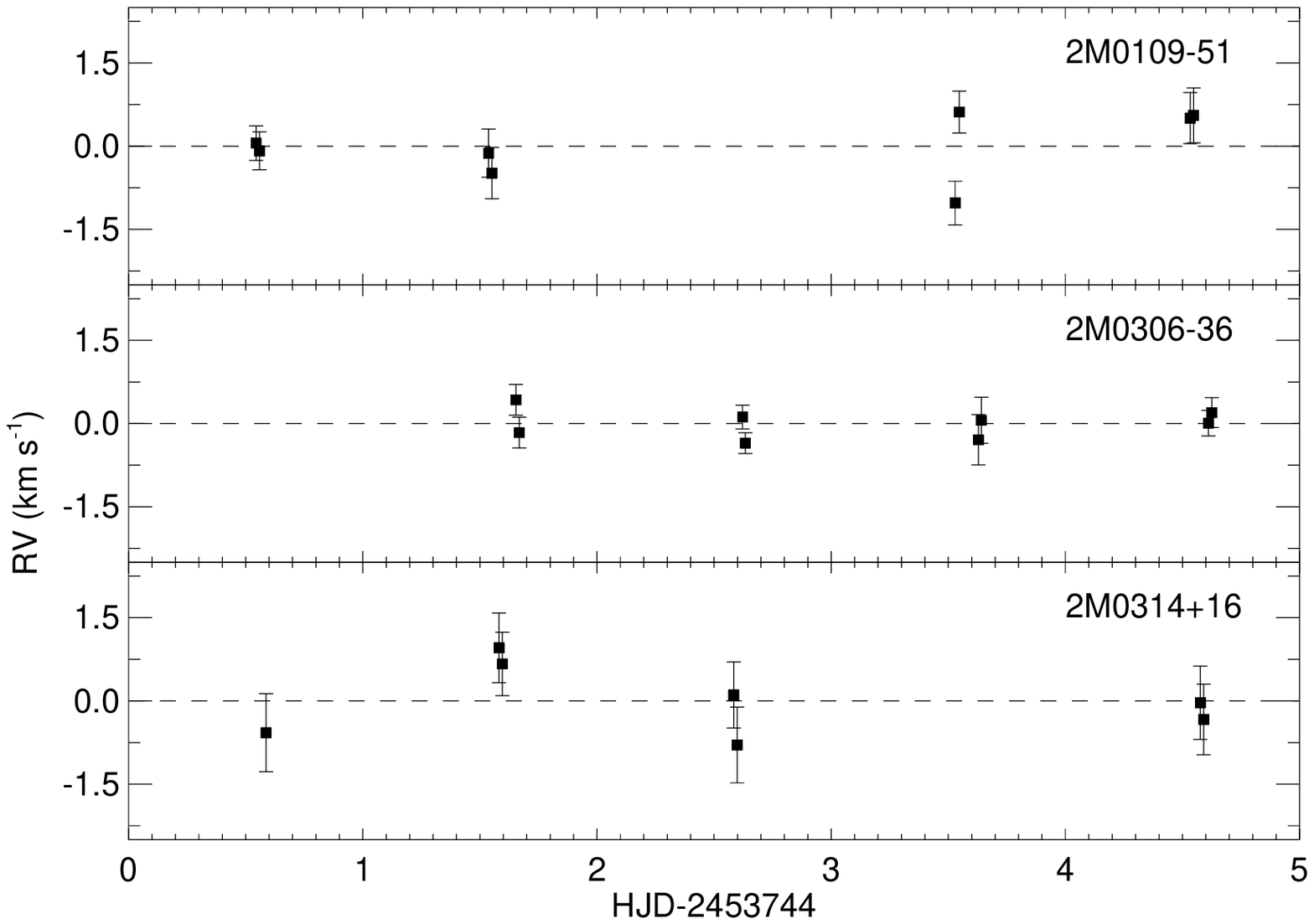}
\caption{Radial-velocity measurements from each position of nodded pairs of spectra derived from the modeling processes described in Section \ref{modeling}. The systemic velocity $V_{\gamma}$ has been subtracted. A barycentric correction has been applied and the errors are estimated from the results of the Monte-Carlo simulations summarized in Figure 4.}
\end{figure}

\begin{figure}
\epsscale{1.0}
\plotone{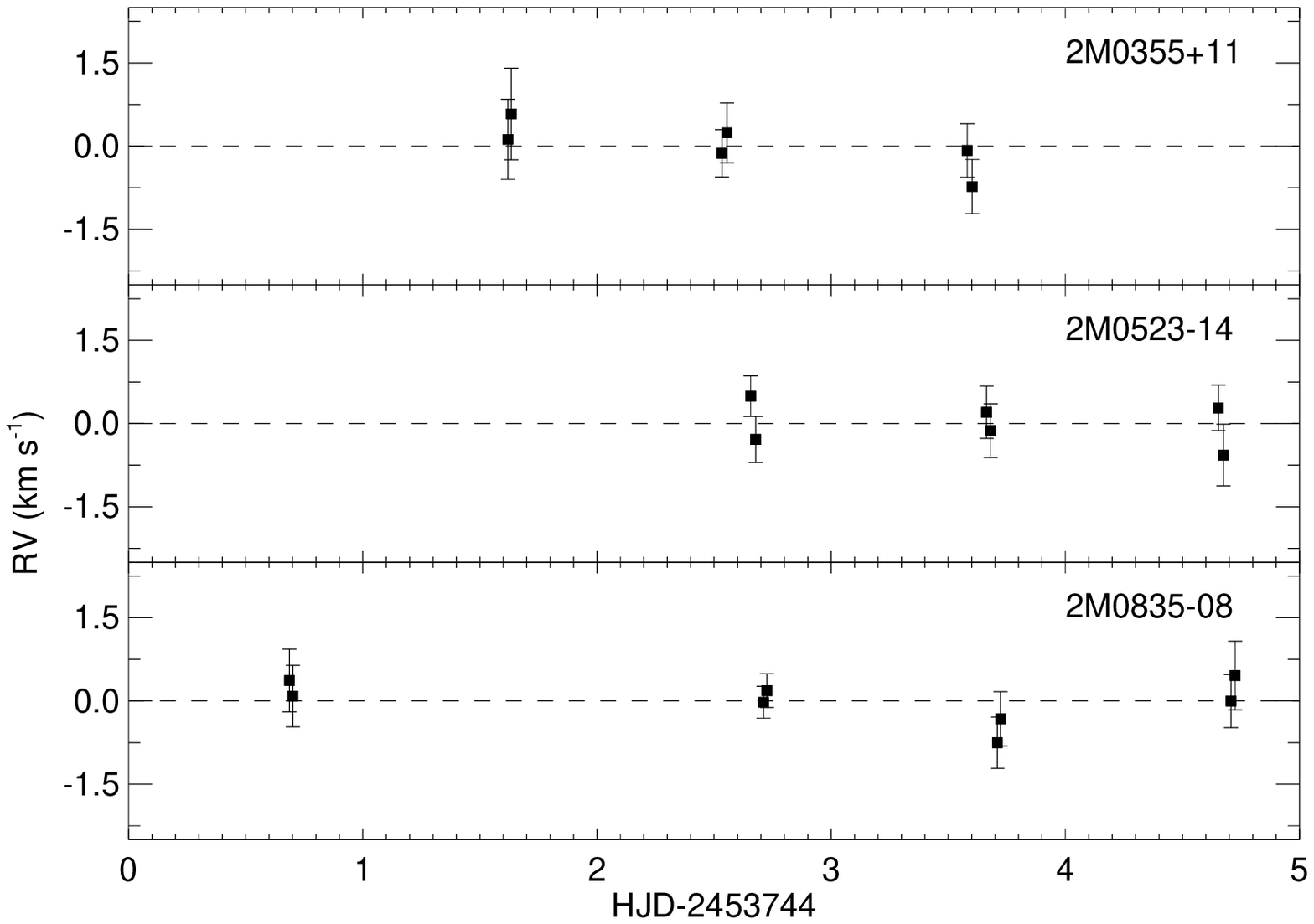}
\caption{Radial-velocity measurements from each position of nodded pairs of spectra derived from the modeling processes described in Section \ref{modeling}. The systemic velocity $V_{\gamma}$ has been subtracted. A barycentric correction has been applied and the errors are estimated from the results of the Monte-Carlo simulations summarized in Figure 4.}
\end{figure}

\begin{figure}
\epsscale{1.0}
\plotone{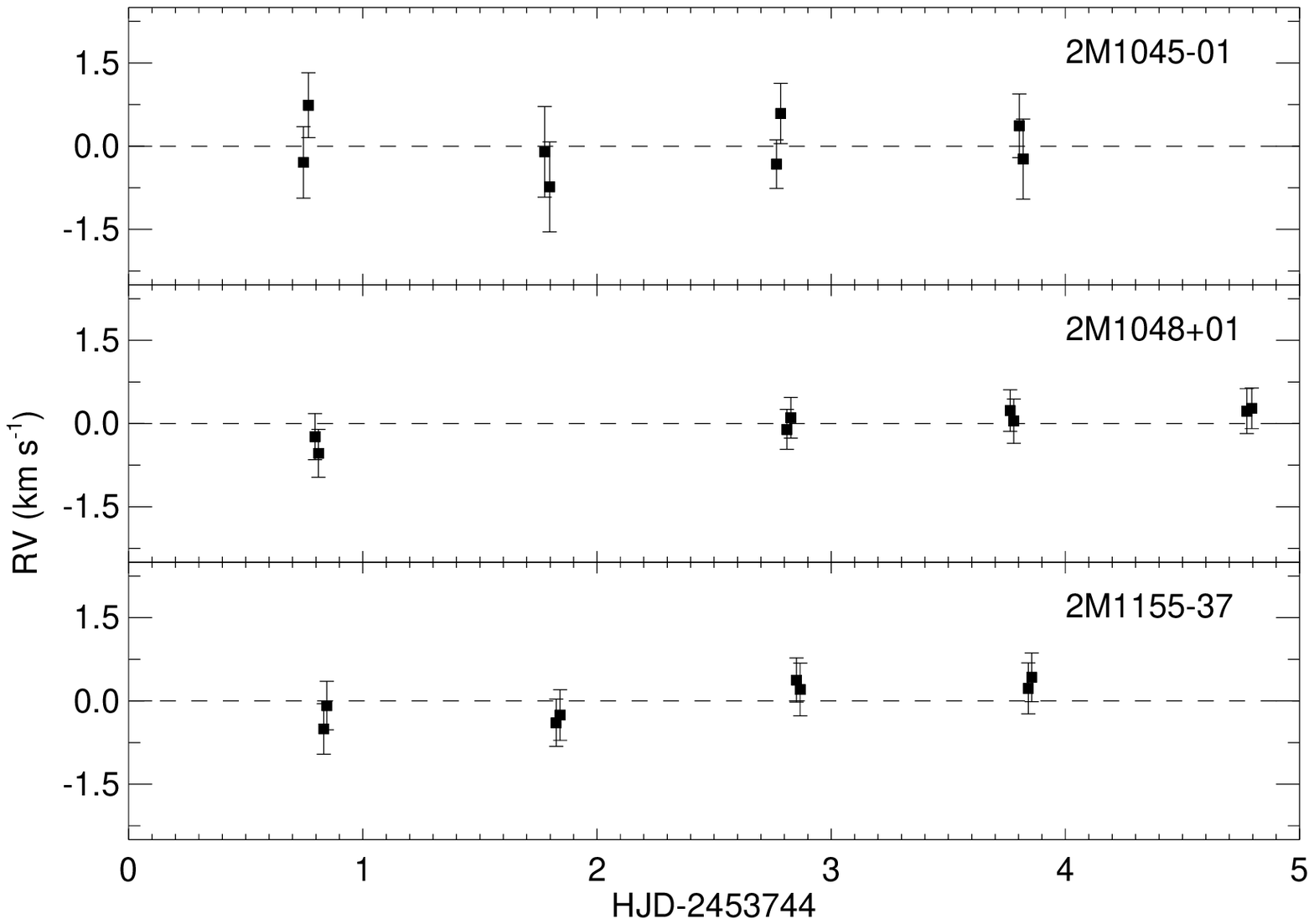}
\caption{Radial velocity measurements from each position of nodded pairs of spectra derived from the modeling processes described in Section \ref{modeling}. The systemic velocity $V_{\gamma}$ has been subtracted.  A barycentric correction has been applied and the errors are estimated from the results of the Monte-Carlo simulations summarized in Figure 4.}
\end{figure}

\clearpage

\begin{deluxetable}{lcccccccccc}
\tabletypesize{\scriptsize}
\tablecaption{Sample}
\tablewidth{0pt}
\tablehead{
\colhead{Object} & \colhead{RA} & \colhead{Dec} & \colhead{K$_{s}$}  & \colhead{Sp. Typ.} & \colhead{Template $T_{\rm{eff}}$} & \colhead{$V\sin i$} & \colhead{V$_{\gamma}$} & \colhead{$\chi_{\nu}^2$} & \colhead{$rms (RV)$} & \colhead{S/N} \\

\colhead{} & \colhead{J2000} & \colhead{J2000} & \colhead{} & \colhead{} & \colhead{K} & \colhead{km s$^{-1}$} & \colhead{km s$^{-1}$} & \colhead{} & \colhead{m s$^{-1}$} & \colhead{} }

\startdata
2M0109--51 & 01:09:02 & --51:00:49 & 11.10 & L2\tablenotemark{a} & 1900  & 14.58$\pm$0.52 & $-$1.30$\pm$0.20 & 1.51 & 570 & 9.6\\

2M0306--36 & 03:06:12 & --36:47:53 & 10.63 & L0\tablenotemark{a} & 1900 & 20.68$\pm$0.65 & +11.58$\pm$0.09 & 1.15 & 260 & 14.2\\

2M0314+16 & 03:14:03 & +16:03:06 & 11.24 & L0\tablenotemark{b} & 2000  & 17.72$\pm$1.32 & $-$6.80$\pm$0.24 & 0.74 & 640 & 5.5\\

2M0355+11 & 03:55:23 & +11:33:44 & 11.53 & L6\tablenotemark{b} & 1900  & 10.31$\pm$0.53 & +12.24$\pm$0.18 & 0.89 & 440 & 7.9\\

2M0523--14 & 05:23:38 & --14:03:02 & 11.64 & L5\tablenotemark{c} & 2100  & 15.26$\pm$0.60 & +11.82$\pm$0.16 & 0.63 & 400 & 8.4\\

2M0835--08 & 08:35:43 & --08:19:24 & 11.14 & L5\tablenotemark{d} & 1600  & 16.92$\pm$0.50 & +29.96$\pm$0.14 & 0.63 & 390 & 9.8\\

2M1045--01 & 10:45:24 & --01:49:58 & 11.78 & L1\tablenotemark{e} & 2200 & 7.06$\pm$0.71 & +6.33$\pm$0.18 & 1.56 & 510 & 7.0\\

2M1048+01 & 10:48:43 & +01:11:58 & 11.62 & L4\tablenotemark{f} & 2200  & 10.29$\pm$0.47 & +24.09$\pm$0.10 & 0.34 & 280 & 8.2\\

2M1155--37 & 11:55:40 & --37:27:35 & 11.46 & L2\tablenotemark{e} & 1600 & 15.93$\pm$0.67 & +45.63$\pm$0.13 & 0.50 & 360 & 8.7\\

\enddata
\tablecomments{The first five columns are object-specific information gathered from dwarfarchives.org. The reported spectral types are based on NIR observations when available, optical observations otherwise. The values of $T_{\rm{eff}}$ represent not an estimate of the actual effective temperature of the object but the temperature of the model spectrum that best fit the data and that was used in the radial velocity analysis. Values of $\chi^2$ are per degree of freedom and represent the fit to the null hypothesis of constant radial velocity. The values of $ rms (RV)$  represent the standard deviation of the radial-velocity measurements for each object treating each nodded position as a separate observation. The values of S/N are the average of the median signal-to-noise per pixel of the individual observations of each object. The errors on $V\sin i$ and V$_{\gamma}$ are statistical and do not take into account likely systematic errors.}
\tablenotetext{a}{\citet{lodieu2005}}
\tablenotetext{b}{\citet{schmidt2007}}
\tablenotetext{c}{\citet{wilson2003}}
\tablenotetext{d}{\citet{cruz2003b}}
\tablenotetext{e}{\citet{gizis2002}}
\tablenotetext{f}{\citet{kendall2004}}

\end{deluxetable}

\begin{deluxetable}{lccc}
\tabletypesize{\scriptsize}
\tablecaption{Radial Velocity Measurements}
\tablewidth{0pt}
\tablehead{
\colhead{Object} & \colhead{HJD-2400000} & \colhead{RV} & \colhead{$\sigma$} \\
\colhead{} & \colhead{} & \colhead{km s$^{-1}$} & \colhead{km s$^{-1}$} }

\startdata
 2M0109$-$51  &  53744.54448  &  $-$1.25  &  0.31\\
 2M0109$-$51  &  53744.55882  &  $-$1.39  &  0.34\\
 2M0109$-$51  &  53745.53691  &  $-$1.43  &  0.43\\
 2M0109$-$51  &  53745.55125  &  $-$1.79  &  0.46\\
 2M0109$-$51  &  53747.52939  &  $-$2.33  &  0.39\\
 2M0109$-$51  &  53747.54720  &  $-$0.69  &  0.38\\
 2M0109$-$51  &  53748.53308  &  $-$0.80  &  0.46\\
 2M0109$-$51  &  53748.54742  &  $-$0.75  &  0.50\\
 2M0306$-$36  &  53745.65385  &  12.01  &  0.28\\
 2M0306$-$36  &  53745.66819  &  11.42  &  0.28\\
 2M0306$-$36  &  53746.62129  &  11.70  &  0.21\\
 2M0306$-$36  &  53746.63330  &  11.23  &  0.19\\
 2M0306$-$36  &  53747.62901  &  11.29  &  0.45\\
 2M0306$-$36  &  53747.64104  &  11.64  &  0.42\\
 2M0306$-$36  &  53748.61104  &  11.59  &  0.23\\
 2M0306$-$36  &  53748.62537  &  11.78  &  0.27\\
 2M0314+16  &  53744.58707  &  $-$7.47  &  0.70\\
 2M0314+16  &  53745.58148  &  $-$5.94  &  0.63\\
 2M0314+16  &  53745.59583  &  $-$6.23  &  0.57\\
 2M0314+16  &  53746.58386  &  $-$6.79  &  0.59\\
 2M0314+16  &  53746.59820  &  $-$7.69  &  0.68\\
 2M0314+16  &  53748.57643  &  $-$6.93  &  0.66\\
 2M0314+16  &  53748.59077  &  $-$7.23  &  0.64\\
 2M0355+11  &  53745.61973  &  12.36  &  0.72\\
 2M0355+11  &  53745.63407  &  12.82  &  0.83\\
 2M0355+11  &  53746.53400  &  12.11  &  0.43\\
 2M0355+11  &  53746.55528  &  12.48  &  0.54\\
 2M0355+11  &  53747.58124  &  12.16  &  0.48\\
 2M0355+11  &  53747.60254  &  11.51  &  0.49\\
 2M0523$-$14  &  53746.65617  &  12.31  &  0.37\\
 2M0523$-$14  &  53746.67747  &  11.53  &  0.42\\
 2M0523$-$14  &  53747.66353  &  12.02  &  0.47\\
 2M0523$-$14  &  53747.68134  &  11.69  &  0.48\\
 2M0523$-$14  &  53748.65352  &  12.10  &  0.41\\
 2M0523$-$14  &  53748.67480  &  11.25  &  0.56\\
 2M0835$-$08  &  53744.68678  &  30.33  &  0.56\\
 2M0835$-$08  &  53744.70112  &  30.05  &  0.55\\
 2M0835$-$08  &  53746.71119  &  29.94  &  0.29\\
 2M0835$-$08  &  53746.72554  &  30.15  &  0.30\\
 2M0835$-$08  &  53747.70936  &  29.21  &  0.46\\
 2M0835$-$08  &  53747.72371  &  29.64  &  0.49\\
 2M0835$-$08  &  53748.70701  &  29.96  &  0.48\\
 2M0835$-$08  &  53748.72484  &  30.42  &  0.62\\
 2M1045$-$01  &  53744.74669  &   6.04  &  0.64\\
 2M1045$-$01  &  53744.76797  &   7.07  &  0.59\\
 2M1045$-$01  &  53745.77659  &   6.23  &  0.82\\
 2M1045$-$01  &  53745.79789  &   5.60  &  0.81\\
 2M1045$-$01  &  53746.76673  &   6.01  &  0.44\\
 2M1045$-$01  &  53746.78455  &   6.92  &  0.54\\
 2M1045$-$01  &  53747.80330  &   6.70  &  0.57\\
 2M1045$-$01  &  53747.81996  &   6.10  &  0.72\\
 2M1048+01  &  53744.79635  &  23.85  &  0.41\\
 2M1048+01  &  53744.81070  &  23.55  &  0.43\\
 2M1048+01  &  53746.81080  &  23.98  &  0.36\\
 2M1048+01  &  53746.82747  &  24.19  &  0.37\\
 2M1048+01  &  53747.76434  &  24.32  &  0.37\\
 2M1048+01  &  53747.77984  &  24.13  &  0.40\\
 2M1048+01  &  53748.77484  &  24.31  &  0.40\\
 2M1048+01  &  53748.79613  &  24.36  &  0.37\\
 2M1155$-$37  &  53744.83327  &  45.13  &  0.46\\
 2M1155$-$37  &  53744.84647  &  45.55  &  0.44\\
 2M1155$-$37  &  53745.82581  &  45.24  &  0.43\\
 2M1155$-$37  &  53745.84248  &  45.38  &  0.46\\
 2M1155$-$37  &  53746.85162  &  46.01  &  0.40\\
 2M1155$-$37  &  53746.86773  &  45.84  &  0.47\\
 2M1155$-$37  &  53747.84184  &  45.86  &  0.46\\
 2M1155$-$37  &  53747.85617  &  46.06  &  0.44\\

\enddata
\tablecomments{Individual radial-velocity observations. Barycentric corrections have been applied to the listed here. The errors on each measurement are estimated  from  the Monte-Carlo simulations described in Section \ref{modeling}.}
\end{deluxetable}

\end{document}